\documentclass[
twocolumn,
]{ceurart}

\usepackage{listings}
\begin{document}

\copyrightyear{2023}
\copyrightclause{Copyright © 2023 for this paper by its authors. Use permitted under Creative Commons License Attribution 4.0 International (CC BY 4.0).}

\conference{Trusting Decentralised Knowledge Graphs and Web Data Workshop,
at European Semantic Web Conference,
May 28 - June 1 in Hersonissos, Greece}

\title{Long-living Service for Cooperative Knowledge Use in Decentralized Data Stores}


\author[1]{Rui Zhao}[%
orcid=0000-0003-2993-2023,
email=rui.zhao@cs.ox.ac.uk,
url=https://renyuneyun.github.io/,
]
\address[1]{University of Oxford,
  Oxford, OX1 2JD, United Kingdom}

\author[1]{Jun Zhao}[%
orcid=0000-0001-6935-9028,
email=jun.zhao@cs.ox.ac.uk,
]

\author[2]{Zimeng Zhou}[%
email=Z.Zhou-45@sms.ed.ac.uk,
]
\address[2]{University of Edinburgh,
  Edinburgh, EH8 9YL, United Kingdom}


\begin{abstract}
  Personal Data Stores (PDS) like SoLiD is an emerging data and knowledge management solution in recent years. They promise to give back ownership and control of data to the user, and provide protocols for developers to build applications using the data. However, existing Solid-based applications often focus on using a single-user's data. In this article, we use a simple but realistic calendar-and-meeting-scheduling scenario to demonstrate the feasibility and design considerations for enabling cooperative data-use across multiple users' SoLiD Pods. This scenario identifies the bottleneck for certain cooperative use cases, namely those involving offline-changing and synchronization of knowledge information. We demonstrate a viable approach to mediate this issue, introducing a long-living thin service, the orchestrator. We describe our implementation and discuss its applicability to other ecosystems. We conclude by discussing the implication of such services, in particular their risks and challenges for building decentralised applications.
\end{abstract}

\begin{keywords}
  Decentralization \sep
  SoLiD  \sep
  Personal Data Store \sep
  Linked Data
\end{keywords}

\maketitle

\section{Introduction}

With the increasing awareness of the importance and value of personal data, legal and technological solutions are being pushed forward for supporting such concerns of the individuals, such as the EU GDPR \cite{noauthor_general_2016} (in particular those related to data portability) and Personal Data Stores (PDS). PDSs like SoLiD \cite{sambra_solid_2016}, openPDS \cite{de_montjoye_openpds_2014} and Databox \cite{mortier_personal_2016} are a promising solution in not only keeping the personal data under individual's control, but also allows the use of such data from other parties to produce values for the individual. In particular, SoLiD is a PDS solution that stores information in Linked Data, which essentially organizes personal and interpersonal data as knowledge graph; it also provides standard Web-compatible protocols for building applications to utilize such data.


However, research and applications (\cite{perera_valorising_2017,zhao_privacy-preserving_2020}) of such systems often focus on using a single store's data, due to the nature of ``personal'' and thus privacy in such systems. There have been limited explorations on cooperative use of data across stores, which can provide exciting opportunities, e.g., for improving our response to global pandemic through aggregating individuals' health and mobility logs\cite{troncoso_decentralized_2020}, to climate change through the sharing of energy consumption habits\cite{baeck_using_2020}, or to improve working condition through analysing many `gig' workers' individual pay rates\cite{zhang_algorithmic_2022}.

We argue that the realization of cooperative data use raises not only privacy challenges but also a series of practical design challenges. In this paper, we will use a simple case -- calendar and meeting scheduling, to illustrate the challenges and design considerations involved in building the cooperative knowledge use application in SoLiD involving interoperability with existing services.
We show how to use a long-living thin service, the orchestrator, to overcome this issue, and our technological solution. Our case study provides important lessons for us to discuss similar challenges to be faced by other cooperative data use applications and what this may imply for the SoLiD ecosystem and beyond. We conclude by discussing the challenges and opportunity of such long-living services for decentralized knowledge use.

\section{Example case -- calendar and meeting scheduling}

Calendar data is an interesting combination of sensitive and non-sensitive data. On one hand, they contain the activities a person would have everyday, which can often be very private; on the other hand, we often want to schedule events (especially meetings) with others by finding joint availability of time slots, which requires sharing some information from everyone's calendar. Existing calendar services often support this requirement by allowing the sharing of the original and the \emph{busy-or-free} projection of the calendar.

The situation will be similar to store calendar data in PDS. But one important advantage of storing them in PDS is the ability to use the calendar data in different ways that were previously unsupported by the existing centralized, or platform-provided services.

\subsection{Work pattern for centralized services}

To find the joint availability, one person needs to be the active party to find the availabilities (and make the decision), and others receive the decision, assuming the calendar information already exists. We call the active party the \emph{activist} while the rest the \emph{passivists}.

In existing centralized calendar systems, to find the joint availability, the users need to:
\begin{enumerate}
    \item (Passivists) Share all calendars to each other user who may schedule meetings with them;
    \item (Activist) Import others' calendars shared to them;
    \item (Activist) Update their local cache (of everyone's calendar) to the latest;
    \item (Activist) Look at the calendars and find the joint availability;
    \item (Activist) Sends the meeting information to all passivists;
    \item (Passivists) Update their calendar with the received information.
\end{enumerate}
It is worth noting that this list provides an extensive view on relevant actions, where in practice not all steps need to be repeated every time (i.e.~Step 1 \& 2) and some steps are usually automated (i.e.~Step 3, 5 \& 6). In particular, Step 5 and 6 are critical for meeting scheduling if multiple activists want to schedule meetings while the passivist is offline, otherwise there can be a clash in slot selection by the different activists.

In practice, Step 1 and 2 both involve manual actions and can be boring and time-costing simultaneously, if the user has many calendars and there are many participants; it will also need to be repeated when a new participant joins.
Some (third-party) services exist (e.g.~Doodle\footnote{Doodle: \url{https://doodle.com/}}) to simplify this process. Their working model mainly involves requiring every user to deposit their calendars to this service, and share or compute internally the time slots. Basically, this keeps a local (to their service) copy of the original calendars and maintain sharing and permission within the (third-party) service itself. This poses a potential privacy concern, as well as a redundancy of data.

\subsection{Support and burden in SoLiD}

Assume that a Solid Calendar App exists that provides individuals to manage their calendar data stored in their data store (i.e.~Pod). In this context, most steps can be mirrored, where the activists and passivists are the Pod owners, and sharing of calendar is the sharing of the calendar resource. An advantage for SoLiD storing data in Linked Data format is that one can organize/advertise their calendars together, and share that to others, instead of sharing each calendar separately, given an agreed common schema\footnote{To simplify the discussion, we assume there is one combined calendar information entry for each Pod owner.}. 


However, there is a problem for Step 5 and 6 in SoLiD or other similar systems: where should the activist send the meeting information to, and how can the calendar information be merged to the calendar information in the passivist's Pod while \emph{offline} (i.e.~no App open)?

This is not a consideration in centralized settings, because the central service will always be \emph{online} (or no service) and can receive and operate on behalf of the passivists. Therefore, this \emph{online} service can handle the merging of incoming meeting information with the calendar of the passivists, even when they are offline. But this does not hold in SoLiD or other decentralised data architectures, which are designed with general mechanism as data and information stores, not specialized for calendar actions.

The mechanism in SoLiD for dealing with incoming information is to use the inbox, a dedicated location for others to write/append information to, implementing the Linked Data Notification (LDN) specification \cite{w3c_linked_2017}. However, it is a generic mechanism -- the Pod will only store the notification to the inbox, and processing of such information is due to the specification of each App. This routes back to requiring the Pod owner to be ``online'' (having appropriate App opened). Therefore, for the meeting notification, this mechanism does not help when a passivist is offline.

There are naive solutions such as giving write permission to the calendar information directly to the activists, or giving everyone read permission to the inbox. However, they can easily form a bad pattern as there are many potential activists, which may result in malicious or incorrect writing due to various reasons, or reading sensitive notifications in the inbox from other applications. Trusting many users and keeping data privacy and integrity forms a challenge.


The problem becomes more complicated if we also consider a hybrid or transitional scenario, where both existing (centralized) calendar services and decentralized data architectures like SoLiD are used simultaneously. This may be inevitable for the time being because some existing workflow or tools only integrate with the centralized calendar services. This is also the requirement to avoid a cold-start of decentralized data architectures.

In this scenario, one must keep synchronization of the calendar within their Pods and that in external calendars (and gradually migrates to Pod-only scenarios). They give permission and use calendar App as above. But instead of storing the meeting information in their Pods, they must also store the information to the external calendar. This would keep compatibility and solve the cold-start issue simultaneously, but also poses challenges on the \emph{synchronization} mechanism while the user is offline.

Thus, we observe the requirements to enable this cooperative data usage involving updating data in passivists' Pod: 1) enable access control of calendar data; 2) interoperate with existing calendar systems; 3) maintain the calendar in synchronization, even when the passivists are offline. Requirement 1 is already supported by Solid, and the rest needs to be address properly.

\section{A solution using orchestrator}

We propose to use a dedicated long-living thin service, called the orchestrator\footnote{Sande et al proposed the concept of orchestrator as an autonomous agent based on triggers; we borrow the name. However, they did not discuss the \emph{necessity} of orchestrators in depth, and it lacks a direct way of supporting the calendar case discussed in this paper.}, to mitigate this issue.
The orchestrator is dedicated to one main functionality: to fetch and transform (external) calendars into the calendar information stored in the user' Pod.

\subsection{Orchestrator in the hybrid scenario}

At the moment, the main scenario is the hybrid case, where the external calendars are the main source of the calendar information in the Pod; later we will discuss how it works for other scenarios, in particular the SoLiD-only scenario.

With the orchestrator, the users still need to grant relevant permission and share information. In the meantime, the relevant synchronization and transformation jobs are performed by the orchestrator.
In particular, as the external calendars are the main source, the orchestrator fetches the specified external calendars, combines them, and stores them into user's Pod. They are controlled by the configuration specified by the user, stored in their Pod.

When scheduling a meeting, the activist picks the time slot, and sends them through the external centralized calendar services. The orchestrator will later pick this up, and synchronize them into the passivists' Pods. This mechanism can also be extended to the inbox-based scenario, as to be discussed later.

Figure \ref{fig:arch} illustrates the architecture and working pattern for using the orchestrator, based on our current implementation. KNoodle is the relevant calendar and meeting scheduling App; the Orchestrator Configurer is the companion App for registering users to the orchestrator and creating configuration files in the Pod. This works for scenarios even when an actor is doing actions based on information in external calendars and another doing actions based on information in the Pod.

The main advantage is that the write operation is only performed by the orchestrator, so the user only needs to trust it, rather than everyone as in the naive solutions; in the meantime, the orchestrator does not have write permission to external calendars, so it will never pollute the information source even if becoming malicious. If for some reason the orchestrator becomes malicious or not functioning, the user can simply revoke the permission on its own Pod; de-registering on the orchestrator is not indispensable for the user to block access from the orchestrator.

Because the orchestrator has a clear role and functionality, it is easy for others to provide alternative orchestrator services for competition, or the calendar owner to self-host. This is the possibility that decentralization and openness brings. The competition may bring multiple benefits, in particular stimulating the orchestrator to provide better and trustworthy service in the long run, and to prevent vendor lock-in of a particular orchestrator provider thus centralization back to the orchestrator provider.

Putting them together, by introducing the orchestrator, the user only needs to trust one agent rather than a large number, controls the critical permission from the Pod, and can switch the orchestrator provider easily. We believe this is a sensible approach with limited drawback to tackle the problem.

\subsection{Implementation with Knoodle}

\begin{figure}
    \centering
    \includegraphics[width=\linewidth]{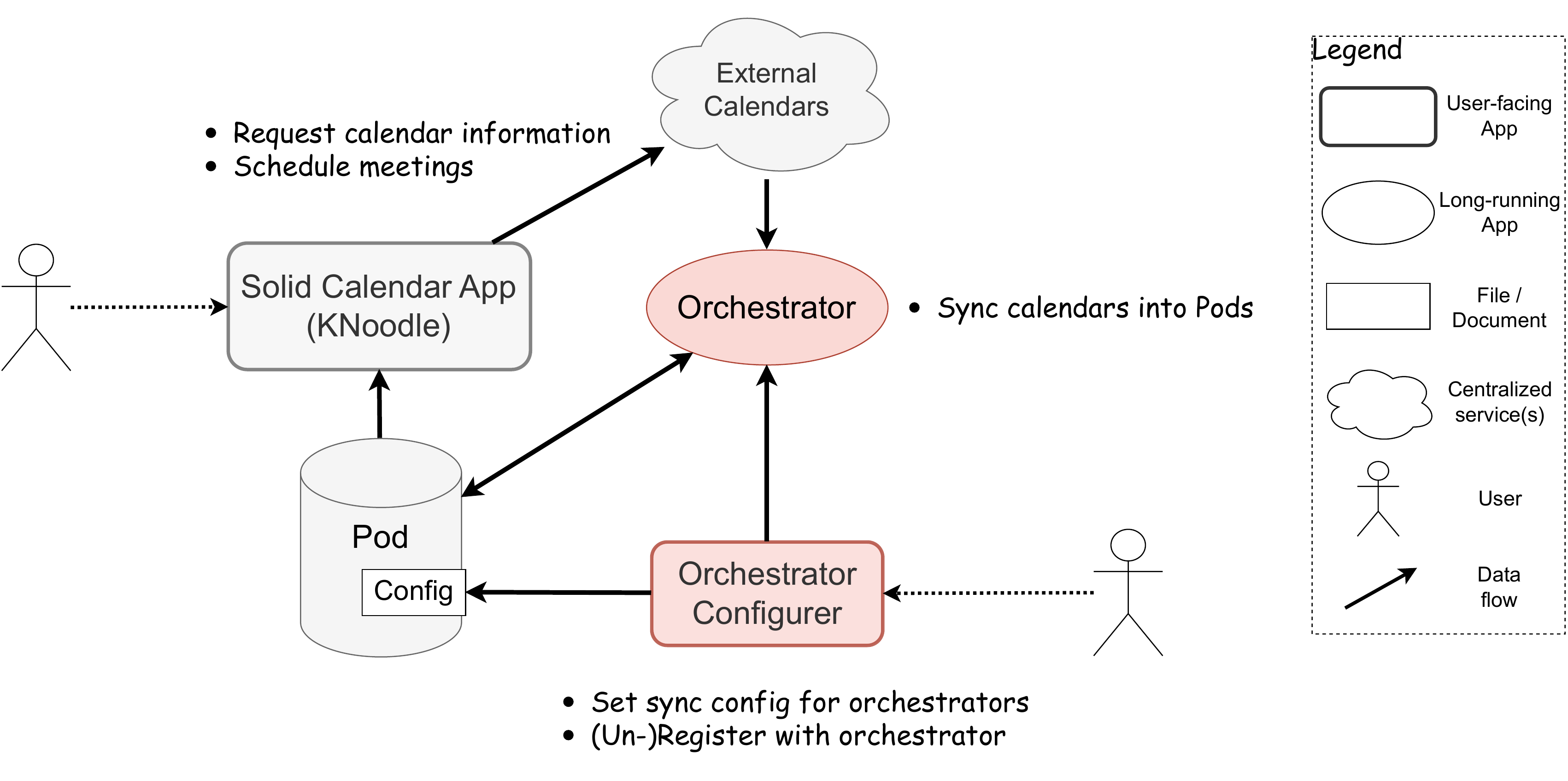}
    \caption{Architecture for the calendar system with the orchestrator.}
    \label{fig:arch}
\end{figure}

We implemented an alternative version of Knoodle\footnote{Knoodle \url{https://github.com/OxfordHCC/knoodle}} with the orchestrator\footnote{Calendar Orchestrator \url{https://github.com/renyuneyun/calendar-orchestrator}}. Knoodle is a SoLiD App for calendar viewing and meeting scheduling, originally developed by the Ghent University. Our version keeps Knoodle as the core user-facing App for meeting scheduling, but also developed the orchestrator to accomplish the full picture, as reflected from Figure \ref{fig:arch}. We also did some functionality and UI improvements, which are omitted in this paper.

In the original version, a modified server implementation (based on Community Solid Server, CSS)\footnote{Solid Calendar Store \url{https://github.com/KNowledgeOnWebScale/solid-calendar-store}} was used. It serves a very similar purpose to the orchestrator, and performs real-time fetching and transformation of external calendars upon request. But it is configured and tightly coupled with the SoLiD service itself. In particular, the user needs to modify the service's configuration file and restart the service to make changes, even if just changing external calendar's URL. The reason it was developed in this way was that it allows some advanced transformations (e.g.~time filtering) and aggregation of external calendar. However, it posed many user experience challenges, particularly with the configuration complexity and portability.

In our version, the orchestrator is deployed as a separate service, not coupled with the SoLiD Pod's service. A companion configuration App is developed to facilitate the user to manage the configuration (stored to user's Pod) and to manage the registration to the orchestrator. The configuration App talks with the orchestrator through its API, thus can be deployed as a normal SoLiD App. Upon registration, the orchestrator obtains an access token from the Pod service, and routinely fetches and updates the calendar information in the Pod. Only the user's basic information (e.g.~WebID, identity provider) and the access token is stored on the orchestrator's local storage. Naturally, one orchestrator can support multiple users simultaneously. Both the orchestrator and the original implementation on modified CSS uses RMLMapper\footnote{RMLMapper \url{https://github.com/RMLio/rmlmapper-java/}} to convert the representation of calendar data.
One can also view the orchestrator implementation as a SoLiD App without user interface\footnote{But contrary to ordinary ephemeral Apps, the orchestrator always runs on the server}.

We are working on improving the orchestrator and configuration App to support more SoLiD implementations, and to support the rich transformation that the Solid Calendar Store supports.

\subsection{Other scenarios}

Although we mainly discussed and designed the orchestrator based on the contemporary situations thus the hybrid scenario, it can be extended/adapted to support other scenarios, in particular the SoLiD-only scenario, without breaking the promises and requirements discussed above.

In a SoLiD-only scenario, the main difference is to use LDN inbox as a replacement of the external calendar service. The orchestrator acts as the \emph{consumer}, reading the calendar notifications in the inbox (in the Pod), and merge that to the calendar information in the Pod. Similarly, the calendar App, as the \emph{sender}, sends the meeting information as notifications to the inbox.

A more complicated scenario is for the SoLiD-first hybrid case, where the information needs to be synchronized for both directions. For the SoLiD-related part, the LDN inbox is required; for the external part, the synchronization mechanism discussed earlier is required. Apart from them, an additional function should be implemented for pushing/synchronizing the new meetings from the LDN inbox to external calendars. Different implementation details (or configuration options) can be provided, such as keeping a separate calendar resource in Pod dedicated for those from inbox and synchronize to external calendar, storing it an ICS calendar in Pod to act as a ``remote'' calendar, or using a separate remote calendar.

In general, the proposed orchestrator mechanism can be extended or adapted for future with the migration progress in decentralized personal stores. It is still essential in those scenarios.

\section{Application to related work}

SoLiD is a good example of decentralized knowledge graph and PDS. It has clear separation of roles, and therefore we anticipate the designs working on SoLiD would also work on other ecosystems as well (with appropriate modification), discussed in this section.

The key observation above is the necessity of long-living services for certain tasks. This needs to be made explicit because SoLiD's PDS service is designed to not run custom code (and only Apps can, with or without UI), and therefore can not be used as the orchestrator.
Some other PDSs also possess the same design, e.g.~myDex\footnote{MyDex \url{https://mydex.org/}} and openPDS \cite{de_montjoye_openpds_2014}. Therefore, the same need of long-living services also exists for them.
Databox \cite{mortier_personal_2016} holds a different view, considering the data and applications as a series of configurable nodes in data flow. In their architecture, it might be possible to develop an application acting as the orchestrator, and install it on every users' store. But it did not present the way to handle events and triggers, still leaving questions.
These other PDSs also lack schemas and can not be treated or queried as a knowledge graph.

In blockchain-based decentralized knowledge graph (e.g.~OriginTrail\footnote{OriginTrail \url{https://origintrail.io/}}), the blockchain itself can store the data and can also execute custom code (in the form of smart contracts), thus reducing the needs of external long-living services -- though reading off-chain data may still require that. However, storing and sharing private or sensitive data on the blockchain remains a challenging task \cite{bernal_bernabe_privacy-preserving_2019}. In addition, the blockchain itself may be a challenge for resources.

\section{Conclusion and open questions}

In this paper, we discussed the problem of supporting calendar-and-meeting-scheduling use case on SoLiD, as an example of cooperative decentralized knowledge use. We showed the problem and proposed the use of long-living service like the orchestrator to support such cases, and presented our implementation and its application to related work.

Bringing up the long-living service like the orchestrator into a decentralized context poses opportunities and threats. On one hand, it is essential for certain tasks for decentralized knowledge use. On the other hand, it poses a potential challenge of re-centralization if the service becomes too big to switch. There is also currently no guarantee that the orchestrator will not be malicious, as they have access to sensitive information either in the configuration or the data obtained. A methodology or mechanism is needed to restrict or mitigate such potentials.
In the meantime, there are also business opportunities on running such services and therefore accelerate adoption of SoLiD or any other decentralized systems.

Apart from investigating the technologies to constraint centralization and improving trust, one may also explore alternative designs that do not involve a third-party long-living service. In particular, one may want to study how to extend the access control model to express the necessary permission for such use cases, thus eliminating the needs.

\begin{acknowledgments}
Special thanks to Pieter Heyvaert from Ghent University for supporting design and implementation.
\end{acknowledgments}

\bibliography{references_shrink,extras}

\end{document}